\documentclass[12pt,a4paper]{article}
\usepackage{mathrsfs}
\usepackage{amsmath}
\usepackage{dsfont}
\usepackage{amsfonts}
\usepackage{amssymb}
\usepackage{graphicx}
\usepackage{txfonts}

\title{Quantum approach to Bertrand duopoly}
\author{
  Piotr Fr\c{a}ckiewicz\\
  \texttt{P.Frackiewicz@impan.gov.pl}
  \and
  Jan S\l adkowski\\
  \texttt{jan.sladkowski@us.edu.pl}
}
\begin{document}
\maketitle
\begin{abstract}
The aim of the paper is to study the Bertrand duopoly example in the quantum domain. We use two ways to write the game in terms of quantum theory. The first one adapts the Li--Du--Massar scheme for the Cournot duopoly. The second one is a simplified model that exploits a two qubit entangled state. In both cases, we focus on finding Nash equilibria in the resulting games. Our analysis allows us to take another look at the classic model of Bertrand.
\end{abstract}
%%%%%%%%%%%%%%%%%%%%%%%%%%%%%%%%%%%%%%%%%
\section{Introduction}
\label{intro}
Quantum game theory is an interdisciplinary field that combines game theory with quantum theory \cite{mejer,ewl,marinatto}. The idea is to use the apparatus developed to describe quantum phenomena to analyze macroscopic complex systems (including living systems) \cite{bel1,khr,bus,hkh,bel2}.  The first attempt to describe a game in the quantum domain applied to finite noncooperative games in the normal form \cite{mejer,ewl,marinatto} but soon after that quantum game theory has found  applications in various fields including decision sciences \cite{khr,bus,imprecall}, finance theory \cite{js5,js1,js2} or mathematical psychology \cite{bus}. Physical implementation of a quantum game could be very hard, and experimental realization of such ideas as quantum auctions \cite{js6,js3} is  a demanding technological challenge, not to mention any commercial use. Fortunately, in some interesting cases no physical creation of entanglement is necessary if one restrict oneself to phenomenological description or modeling of agent's behavior. There are suggestions that quantum games can outperform the "classical" ones in description of some interesting phenomena in economic theory or social sciences\footnote{We should stress here that quantum games are games in the standard sense and the reader should not assign any mysterious contexts to the adjective quantum.} \cite{bus,hkh,js4,aer,yuk,mak}. In this way, quantum game theory has developed into an independent analytical tool that uses the sometimes possible advantage of dealing with probability amplitudes instead of probabilities. The analysis of oligopolies has a long history \cite{puu} and attempts at exploring quantum game theory to this field of research should not surprise us. Most of the attention is focused on the duopoly theory hoping for more reliable modeling. The generally accepted quantum scheme for these problems is due to Li et al. \cite{du}.
At present, one can find papers that apply the Li--Du--Massar scheme to the Cournot duopoly problems \cite{3.2,{4.1},6} and the Stackelberg duopoly \cite{5,{8.1},8.2}. There is also the study of quantum games concerning Bertrand duopoly examples \cite{kiang,qin}. The motivation of writing this paper was twofold. One of the purposes is to extend the quantum game theory based on the Li--Du--Massar scheme so that another type of duopoly has the quantum analogue. The Bertrand duopoly in the form studied in the paper is an alternative model of the famous Cournot duopoly, where the players compete in prices instead of quantities. This change makes it impossible to have a positive equilibrium outcome in the Bertrand duopoly. Thus, it would be interesting to study this problem in the quantum domain.

On the other hand, our goal is to apply a method of determining Nash equilibria that is a new one with respect to the quantum duopolies (whereas it is commonly used in the classical game theory). The Bertrand duopoly example is determined by a piecewise payoff function. As a result, the payoff function in the quantum game has similar form. It requires more sophisticated methods to find all the Nash equilibria than ones appeared in the previous papers.
%%%%%%%%%%%%%%%%%%%%%%%%%%%%%%%%%%%%%%%%%%%%%%%
\section{Bertrand duopoly problem}
Let us recall the classical problem of Bertrand duopoly. There are two firms (players) who compete in the price of a homogenous product. The demand $q$ of the product is a function of the price, $q(p) = \max{\{a-p,0\}}$ for every $p \geqslant 0$.  The firm with a lower price captures the entire market. If both firms charge the same price, they split the market equally. We assume that each firm has the same marginal cost $c$ such that $0\leqslant c < a$. If player 1 sets the price as $p_{1}$ and player 2 sets the price as $p_{2}$ the payoff function of player 1 is
\begin{equation}\label{bertrand1}
u_{1}(p_{1}, p_{2}) = \begin{cases}(p_{1} - c)(a-p_{1}) &\mbox{if}~p_{1} < p_{2}~\mbox{and}~p_{1} \leqslant a, \cr \frac{1}{2}(p_{1} -c)(a-p_{1}) &\mbox{if}~p_{1}=p_{2}~\mbox{and}~p_{1} \leqslant a,\cr 0 &\mbox{if}~\mbox{otherwise}.\end{cases}
\end{equation}
Similarly, the payoff function of player 2 is
\begin{equation}\label{bertrand2}
u_{2}(p_{1}, p_{2}) = \begin{cases}(p_{2} - c)(a-p_{2}) &\mbox{if}~p_{2} < p_{1}~\mbox{and}~p_{2} \leqslant a, \cr \frac{1}{2}(p_{2} -c)(a-p_{2}) &\mbox{if}~p_{1}=p_{2}~\mbox{and}~p_{2} \leqslant a,\cr 0 &\mbox{if}~\mbox{otherwise}.\end{cases}
\end{equation}
The Bertrand model \cite{bertrand} was proposed as an alternative to the Cournot model~\cite{cournot} in which the players compete in quantities (see also \cite{peters} for more details about these two models). While it seems that the rational players would obtain similar payoffs in both games, comparison of the Cournot and Bertrand duopoly examples with respect to Nash equilibria exhibits a paradox. In the Cournot duopoly, the Nash equilibrium payoff is $(a-c)^2/9$. On the other hand, the game defined by~(\ref{bertrand1})--(\ref{bertrand2}) has the unique Nash equilibrium $(p^*_{1}, p^*_{2}) = (c,c)$ that arises from intersection of best reply functions $\beta_{1}(p_{2})$ and $\beta_{2}(p_{1})$,
\begin{eqnarray}\label{classicalbestreply}
\beta_{1}(p_{2}) = \begin{cases}\{p_{1}|p_{1}>p_{2}\} &\mbox{if}~ p_{2} < c, \cr\{p_{1} | p_{1}\geqslant c\} &\mbox{if}~ p_{2} = c, \cr \varnothing &\mbox{if}~ c< p_{2} \leqslant \frac{a+c}{2}, \cr \left\{\frac{a+c}{2}\right\} &\mbox{if}~ p_{2} > \frac{a+c}{2},\end{cases}\\ \label{classicalbestreply2}
\beta_{2}(p_{1}) = \begin{cases}\{p_{2}|p_{2}>p_{1}\} &\mbox{if}~ p_{1} < c, \cr\{p_{2} | p_{2}\geqslant c\} &\mbox{if}~ p_{1} = c, \cr \varnothing &\mbox{if}~ c< p_{1} \leqslant \frac{a+c}{2}, \cr \left\{\frac{a+c}{2}\right\} &\mbox{if}~ p_{1} > \frac{a+c}{2}.\end{cases}
\end{eqnarray}
The equilibrium implies the payoff of 0 for both players.
%%%%%%%%%%%%%%%%%%%%%%%%%%%%%%%%%%%%%%%%%%%%%%%
\section{Quantum Bertrand duopoly}
In \cite{remarks} we discussed two well-known quantum duopoly schemes~\cite{du,iqbalbackwards}. We pointed out that under some condition the Li--Du--Massar scheme \cite{du} appears to be more reasonable. In what follows, we apply that scheme to Bertrand duopoly problem and study the resulting game with respect to Nash equilibria. Next, we investigate the duopoly problem with the use of a simpler two-qubit scheme.
\subsection{The Li--Du--Massar approach to Bertrand duopoly}
Let us recall the key elements of the Li--Du--Massar scheme that are needed to consider the Bertrand duopoly. In the original paper \cite{du}, the quantities $q_{1}$ and $q_{2}$ in the quantum Cournot duopoly are determined by the measurements $\hat{X}_{1}$ and $\hat{X}_{2}$ on the final state $|\Psi_{f}\rangle$. Formally, the final state is of the form:
\begin{equation}
|\Psi_{f}\rangle = \hat{J}(\gamma)^{\dag}(\hat{D}_{1}(x_{1})\otimes \hat{D}_{2}(x_{2}))\hat{J}(\gamma)|0\rangle_{1}|0\rangle_{2},
\end{equation}
where
\begin{itemize}
\item $\hat{J}(\gamma)$ is the entangling operator,
\begin{equation}
\hat{J}(\gamma) = \mathrm{e}^{-\gamma(\hat{a}^{\dag}_{1}\hat{a}^{\dag}_{2} - \hat{a}_{1}\hat{a}_{2})},
\end{equation}
\item $\hat{D}_{j}(x_{j})$ for $x_{j} \in [0,\infty)$ and $j=1,2$ are unitary operators
\begin{equation}
\hat{D}_{j}(x_{j}) = \mathrm{e}^{x_{j}(\hat{a}^{\dag}_{j}-\hat{a}_{j})/\sqrt{2}}
\end{equation}
that correspond to player $j$'s strategies,
\item operators $\hat{a}_{j}$ and $\hat{a}^{\dag}_{j}$ satisfy the following commutation relations:
\begin{equation}\label{commutation}
[\hat{a}_{i}, \hat{a}^{\dag}_{j}] = \delta_{ij}, ~[\hat{a}^{\dag}_{i}, \hat{a}^{\dag}_{j}] = [\hat{a}_{i}, \hat{a}_{j}] = 0.
\end{equation}
\end{itemize}
Then the quantities $q_{1}$ and $q_{2}$ are obtained by formula
\begin{eqnarray}\label{wynik1}
&&q_{1}\equiv \langle \Psi_{f}|\hat{X}_{1}|\Psi_{f}\rangle = x_{1}\cosh{\gamma} + x_{2}\sinh{\gamma},\\\label{wynik2}
&&q_{2}\equiv \langle \Psi_{f}|\hat{X}_{2}|\Psi_{f}\rangle = x_{2}\cosh{\gamma} + x_{1}\sinh{\gamma}.
\end{eqnarray}
In what follows, we provide the reader with detailed calculation needed to obtain (\ref{wynik1}). The same reasoning applies to the case~(\ref{wynik2}).

First, we recall the following operator relation that involves the function $\mathrm{e}^{A}$ (see also \cite{eugen}):
\begin{equation}\label{eidentity}
\mathrm{e}^{\lambda A}B\mathrm{e}^{-\lambda A} = B\cosh{\gamma\sqrt{\beta}} + \frac{[A,B]}{\sqrt{\beta}}\sinh{\gamma \sqrt{\beta}}
\end{equation}
for operators $A$ and $B$ that satisfy $[A, [A,B]] = \beta B$, ($\beta$: constant). In a special case $[A,B] = \mu\mathbb{1}$ ($\mu$: constant), formula~(\ref{eidentity}) leads to
\begin{equation}\label{uproszczonyeidentity}
\mathrm{e}^{\lambda A}B\mathrm{e}^{-\lambda A} = B + \lambda \mu \mathbb{1}.
\end{equation}
From (\ref{commutation}) and (\ref{eidentity}) we have
\begin{equation}
\hat{J}(\gamma)\hat{a}_{1}\hat{J}^{\dag}(\gamma) = \hat{a}_{1}\cosh{\gamma} + \hat{a}^{\dag}_{2}\sinh{\gamma}.
\end{equation}
Thus, we have
\begin{equation}
\hat{O}_{1} \equiv \hat{J}(\gamma)\hat{X}_{1}\hat{J}^{\dag}(\gamma) = \hat{X}_{1}\cosh{\gamma} + \hat{X}_{2}\sinh{\gamma}.
\end{equation}
Applying~(\ref{uproszczonyeidentity}), we obtain
\begin{equation}
\hat{D}^{\dag}_{i}(x_{i})\hat{a}_{j}\hat{D}_{i}(x_{i}) = \begin{cases}\hat{a}_{i} + \frac{x_{i}}{\sqrt{2}} &\mbox{if}~i=j, \cr
\hat{a}_{j} &\mbox{if}~i\ne j,\end{cases}
\end{equation}
for $i,j = 1,2$. Therefore
\begin{equation}
\hat{O}_{2} \equiv \hat{D}^{\dag}_{2}(x_{2})\hat{D}^{\dag}_{1}(x_{1})\hat{O}_{1}\hat{D}_{1}(x_{1})\hat{D}_{2}(x_{2}) = (\hat{X}_{1} + x_{1})\cosh{\gamma} + (\hat{X}_{2} + x_{2})\sinh{\gamma}.
\end{equation}
Since,
\begin{equation}
\hat{J}^{\dag}(\gamma)\hat{a}_{i}\hat{J}(\gamma) = \hat{a}_{i}\cosh{\gamma} - \hat{a}^{\dag}_{j}\sinh{\gamma},
\end{equation}
for $i, j = 1,2$ and $i\ne j$, we thus get
\begin{equation}
\hat{O}_{3} \equiv \hat{J}^{\dag}(\gamma)\hat{O}_{2}\hat{J}(\gamma) = \hat{X}_{1} + x_{1}\cosh{\gamma} + x_{2}\sinh{\gamma}.
\end{equation}
According to the theory of quantization of the electromagnetic field, operators $\hat{a}_{i}$ and $\hat{a}^{\dag}_{i}$ satisfy relations
\begin{equation}
\hat{a}_{i}|n\rangle = \sqrt{n}|n-1\rangle, ~\hat{a}^{\dag}_{i}|n\rangle = \sqrt{n+1}|n+1\rangle.
\end{equation}
Hence
\begin{equation}
\langle 0|_{1} \hat{O}_{3}|0\rangle_{1} = x_{1}\cosh{\gamma} + x_{2}\sinh{\gamma}.
\end{equation}

We now apply the Li--Du--Massar scheme to the Bertrand duopoly example.
%%%%%%%%%%%%%%%%%%%%%%%%%%%%%%%%%%%%%%%%%%%%%%
From a~game-theoretical point of view, the players 1 and 2 are to choose $x_{1}, x_{2} \in [0,\infty)$, respectively. Then, the players' prices $p_{1}$ and $p_{2}$ are determined as functions $p_{i}\colon [0,\infty)^3 \to [0,\infty)$ of $x_{1}, x_{2}$ and a fixed entanglement parameter $\gamma \in [0,\infty)$,
\begin{equation}\label{quantumprice}
\begin{cases}p_{1}(x_{1},x_{2}, \gamma) = x_{1}\cosh{\gamma} + x_{2}\sinh{\gamma}, \cr p_{2}(x_{1},x_{2}, \gamma) = x_{2}\cosh{\gamma} + x_{1}\sinh{\gamma},\end{cases}
\end{equation}
(see \cite{du} and the papers \cite{kiang,{qin},sekiquchi} directly related to Bertrand duopoly-type problems for justifying formula~(\ref{quantumprice}) in terms of quantum theory). Substituting~(\ref{quantumprice}) into (\ref{bertrand1}) and (\ref{bertrand2}) and noting that the sign of $p_{1}(x_{1}, x_{2}, \gamma) - p_{2}(x_{1}, x_{2}, \gamma)$ depends on the sign of $x_{2} -x_{1}$ we obtain the following quantum counterpart of (\ref{bertrand1}) and (\ref{bertrand2}):
\begin{eqnarray}\label{quantumu}
&&u^{Q}_{1}(x_{1}, x_{2}) = \begin{cases}(p_{1}(x_{1},x_{2}, \gamma) - c)(a - p_{1}(x_{1},x_{2}, \gamma)) &\mbox{if}~ x_{1} < x_{2}, p_{1}(x_{1},x_{2}, \gamma) \leqslant a, \cr \frac{1}{2}(p_{1}(x_{1},x_{2}, \gamma) - c)(a - p_{1}(x_{1},x_{2}, \gamma)) &\mbox{if}~ x_{1} = x_{2}, x_{1}\mathrm{e}^{\gamma} \leqslant a, \cr 0 &\mbox{if}~ \mbox{otherwise},\end{cases}\\ \label{quantumu2}
&&u^{Q}_{2}(x_{1}, x_{2}) = \begin{cases}(p_{2}(x_{1},x_{2}, \gamma) - c)(a - p_{2}(x_{1},x_{2}, \gamma)) &\mbox{if}~ x_{2} < x_{1}, p_{2}(x_{1},x_{2}, \gamma) \leqslant a, \cr \frac{1}{2}(p_{2}(x_{1},x_{2}, \gamma) - c)(a - p_{2}(x_{1},x_{2}, \gamma)) &\mbox{if}~ x_{1} = x_{2}, x_{2}\mathrm{e}^{\gamma} \leqslant a, \cr 0 &\mbox{if}~ \mbox{otherwise}.\end{cases}
\end{eqnarray}
%%%%%%%%%%%%%%%%%%%%%%%%%%%%%%%%%%%%%%%%%%%%%%
\subsubsection*{Nash equilibrium analysis}
In order to find all the Nash equilibria, we determine the best reply functions $\beta_{1}(x_{2})$ and $\beta_{2}(x_{1})$ and find the points of intersection of the graphs of these functions. For $\gamma = 0$ we have $p_{1}(x_{1},x_{2}, 0) = x_{1}$ and $p_{2}(x_{1},x_{2}, 0) = x_{2}$. Then $\beta_{1}(x_{2})$ and $\beta_{2}(x_{1})$ coincide with the classical best reply functions (\ref{classicalbestreply}) and (\ref{classicalbestreply2}). We thus assume that $\gamma > 0$.

Let us consider several cases to settle $\beta_{1}(x_{2})$.
\begin{enumerate}
\item If $x_{2} < c/\mathrm{e}^{\gamma}$, player 1 obtains a negative payoff by choosing $x_{1} \leqslant x_{2}$. Indeed,
\begin{equation}
x_{1}\cosh{\gamma} + x_{2}\sinh{\gamma} - c < \frac{c}{\mathrm{e}^{\gamma}}\cosh{\gamma} + \frac{c}{\mathrm{e}^{\gamma}}\sinh{\gamma} - c = 0.
\end{equation}
and $a - (x_{1}\cosh{\gamma} + x_{2}\sinh{\gamma}) > a -c > 0$. Hence, according to~(\ref{quantumu}), it is optimal for player 1 to take $x_{1} > x_{2}$. By a similar argument, if $x_{2} = c/\mathrm{e}^{\gamma}$, then any $x_{1} < x_{2}$ yields player 1 a negative payoff. For this reason, player 1's best reply is $x_{1} \geqslant c/\mathrm{e}^{\gamma}$. In that case, player 1 obtains the payoff of 0.

\item Let us now consider the case $c/\mathrm{e}^{\gamma} < x_{2} \leqslant (a+c)/(2\mathrm{e}^{\gamma})$. Note that
\begin{equation}\label{haha}
p_{1}(x_{1}, x_{2}, \gamma) = x_{1}\cosh{\gamma} + x_{2}\sinh{\gamma} = (a+c)/2
\end{equation}
maximizes expression
\begin{eqnarray}\label{expression}
&&(p_{1}(x_{1},x_{2},\gamma) - c)(a-p_{1}(x_{1},x_{2},\gamma)) \nonumber\\ &&\quad = (x_{1}\cosh{\gamma} + x_{2}\sinh{\gamma} -c)(a - (x_{1}\cosh{\gamma} + x_{2}\sinh{\gamma})).
\end{eqnarray}
Hence, if $x_{2} = (a+c)/(2\mathrm{e}^{\gamma})$, term~(\ref{expression}) as a function of variable $x_{1}$ is maximized at $x_{1} = (a+c)/(2\mathrm{e}^{\gamma})$. However, equality $x_{1} = x_{2}$ implies that the players split the payoff given by~(\ref{expression}). Thus, player 1 would benefit from choosing $x_{1}$ slightly below $x_{2}$. But then any $x'_{1}$ in between $x_{1}$ and $x_{2}$ would yield a better payoff. As a result, there is no best reply in this case. If $c/\mathrm{e}^{\gamma} < x_{2} < (a+c)/(2\mathrm{e}^{\gamma})$ then it follows from~(\ref{haha}) that expression~(\ref{expression}) is maximized at point $x_{1} > (a+c)/(2\mathrm{e}^{\gamma}) > x_{2}$. But by taking into account payoff function~(\ref{quantumu}), it would result in player 1's payoff of 0. Thus, player 1 again obtains more by choosing $x_{1}$ slightly below $x_{2}$. In the same manner as in case $x_{2} =  (a+c)/(2\mathrm{e}^{\gamma})$ we can see that the set of best responses of player 1 is empty when $x_{2} < (a+c)/(2\mathrm{e}^{\gamma})$.

\item If $(a+c)/(2\mathrm{e}^{\gamma}) < x_{2} \leqslant (a + c)/(2\sinh{\gamma})$, then from the fact that $x_{1}\cosh{\gamma} + x_{2}\sinh{\gamma} = (a+c)/2$ maximizes~(\ref{expression}) the player 1's best reply is $x_{1} = ((a+c)/2 - x_{2}\sinh{\gamma})/\cosh{\gamma}$

\item If $(a+c)/(2\sinh{\gamma}) < x_{2} < a/\sinh{\gamma}$, function~(\ref{expression}) of variable $x_{1}$ is monotonically decreasing in interval $[0,\infty)$. Hence, player 1 would obtain the highest payoff if $x_{1} = 0$.

\item For the case $x_{2} \geqslant a/\sinh{\gamma}$ we have $p_{1}(x_{1}, x_{2}, \gamma)\geqslant a$ for any $x_{1} \in [0,\infty)$. It follows that $u^{Q}_{1}(x_{1},x_{2}) = 0$, and then the set of best replies is $[0,\infty)$.
\end{enumerate}
Summarizing, we obtain the following best reply function $\beta_{1}(x_{2})$:
\begin{equation}\label{bestreplyq1}
\beta_{1}(x_{2}) = \begin{cases}\{x_{1}\colon x_{1}>x_{2}\} &\mbox{if}~ x_{2} < \frac{c}{\mathrm{e}^{\gamma}}, \cr \left\{x_{1}\colon x_{1}\geqslant \frac{c}{\mathrm{e}^{\gamma}}\right\} &\mbox{if} x_{2} = \frac{c}{\mathrm{e}^{\gamma}}, \cr \varnothing &\mbox{if}~ \frac{c}{\mathrm{e}^{\gamma}} < x_{2} \leqslant \frac{a+c}{2\mathrm{e}^{\gamma}}, \cr \left(\frac{a+c}{2} - x_{2}\sinh{\gamma}\right)\mathrm{sech}{\,\gamma} &\mbox{if}~ \frac{a+c}{2\mathrm{e}^{\gamma}} < x_{2} \leqslant \frac{a+c}{2\sinh{\gamma}}, \cr 0 &\mbox{if} \frac{a+c}{2\sinh{\gamma}} < x_{2} < \frac{a}{\sinh{\gamma}}, \cr [0,\infty) &\mbox{if}~ x_{2} \geqslant \frac{a}{\sinh{\gamma}}. \end{cases}
\end{equation}
%tutaj skończyłem%
Similar arguments to those above show that player 2's best reply function $\beta_{2}(x_{1})$ is
\begin{equation}\label{bestreplyq2}
\beta_{2}(x_{1}) = \begin{cases}\{x_{2}\colon x_{2}>x_{1}\} &\mbox{if}~ x_{1} < \frac{c}{\mathrm{e}^{\gamma}}, \cr \left\{x_{2}\colon x_{2}\geqslant \frac{c}{\mathrm{e}^{\gamma}}\right\} &\mbox{if}~ x_{1} = \frac{c}{\mathrm{e}^{\gamma}}, \cr \varnothing &\mbox{if}~ \frac{c}{\mathrm{e}^{\gamma}} < x_{1} \leqslant \frac{a+c}{2\mathrm{e}^{\gamma}}, \cr \left(\frac{a+c}{2} - x_{1}\sinh{\gamma}\right)\mathrm{sech}{\,\gamma} &\mbox{if}~ \frac{a+c}{2\mathrm{e}^{\gamma}} < x_{1} \leqslant \frac{a+c}{2\sinh{\gamma}}, \cr 0 &\mbox{if}~ \frac{a+c}{2\sinh{\gamma}} < x_{1} < \frac{a}{\sinh{\gamma}} \cr [0,\infty) &\mbox{if}~ x_{1} \geqslant \frac{a}{\sinh{\gamma}}. \end{cases}
\end{equation}
It is clear now that the players best reply functions~$\beta_{1}(x_{2})$, $\beta_{2}(x_{1})$ for $\gamma \ne 0$ are more complex compared with (\ref{classicalbestreply}). If $\gamma \ne 0,$ the best reply functions on interval $((a+c)/(2\mathrm{e}^{\gamma}), \infty)$ (being the counterpart of case $x_{i} > \frac{a+c}{2}$ if $\gamma =0$) are more specified, and take into account different intervals $((a+c)/(2\mathrm{e}^{\gamma}), (a+c)/(2\sinh{\gamma})]$, $((a+c)/(2\sinh{\gamma}), a/\sinh{\gamma})$ and $[a/\sinh{\gamma}, \infty)$. This implies that new equilibria arise. Since a Nash equilibrium in a two-person game is a strategy profile in which the strategies are mutually best replies, we can easily determine the Nash equilibria by studying the points of intersection of $\beta_{1}(x_{2})$ and $\beta_{2}(x_{1})$ (see Fig~1 for the graphs of $\beta_{1}(x_{2})$ and $\beta_{2}(x_{1})$).
%%%%%%%%%%%%%%%%%%%%%%%%%%%%%%%%%%%%%%%%%%%%%%
\begin{figure}[t]
\includegraphics[scale=0.6]{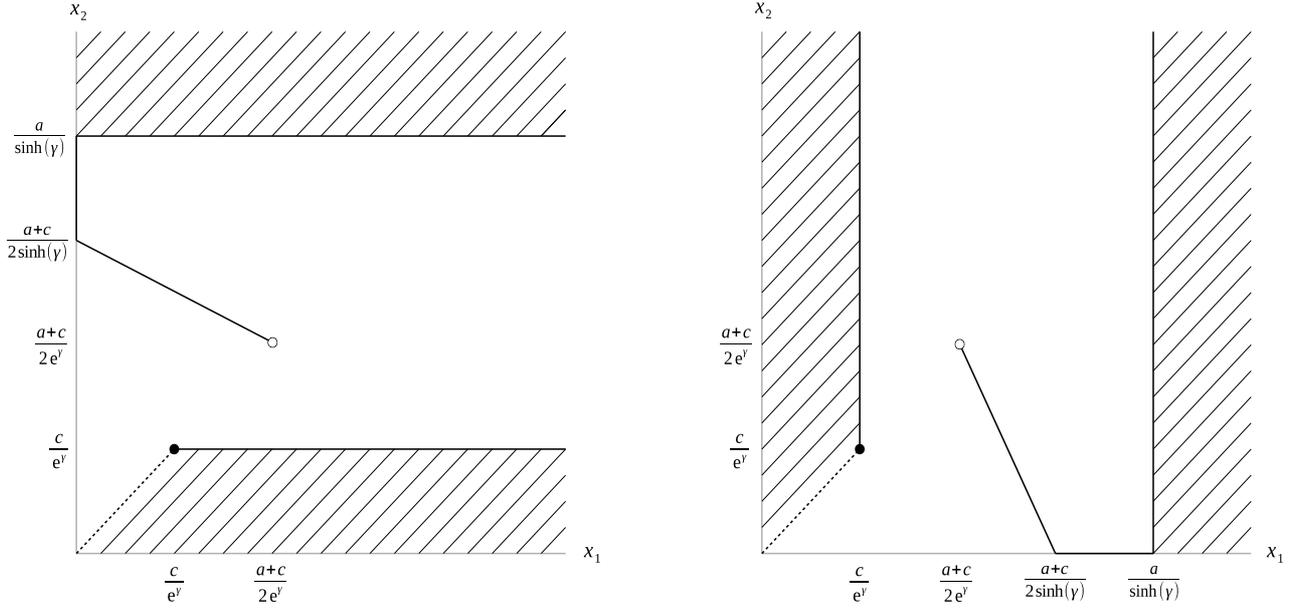}
%\resizebox*{0.72\textwidth}{0.36\textheight}{\includegraphics[scale=1]{figpay6_111_w_ghz.jpg}}
\caption{Graphs of best reply functions (\ref{bestreplyq1}) and (\ref{bestreplyq2}). \label{figure1}}
\end{figure}
%%%%%%%%%%%%%%%%%%%%%%%%%%%%%%%%%%%%%
An example of such a point is profile $(c/\mathrm{e}^{\gamma}, c/\mathrm{e}^{\gamma})$ that coincides with the unique classical Nash equilibrium $(c, c)$ in case $\gamma = 0$. Another and more interesting example is a profile $(0, (a+c)/(2\sinh{\gamma}))$ that implies the payoff profile $((a-c)^2/4, 0)$ and has no counterpart in the classical Cournot duopoly. It is a particular case of a general equilibrium profile 
\begin{equation}
\left(x_{1}, \left(\frac{a+b}{2} - x_{1}\cosh{\gamma}\right)\mathrm{csch}{\gamma}\right)~\mbox{for}~x_{1} \in \left[0, \frac{c}{\mathrm{e}^{\gamma}}\right].
\end{equation}
To see that this type of Nash equilibrium yields the payoff profile $((a-c)^2/4, 0)$, note first that $x_{1} \leqslant c/\mathrm{e}^{\gamma}$ implies $\mathrm{e}^{\gamma}x_{1} < a+c$. From this we conclude
\begin{equation}
x_{1} < \left(\frac{a+c}{2} - x_{1}\cosh{\gamma}\right)\mathrm{csch}{\gamma} = x_{2}.
\end{equation}
As a result, player 2's payoff is equal to zero and player 1's payoff function $u^{Q}_{1}(x_{1}, x_{2})$ comes down to
\begin{equation}
u^{Q}_{1}(x_{1}, x_{2}) = (p_{1}(x_{1}, x_{2}, \gamma) - c)(a - p_{1}(x_{1}, x_{2}, \gamma)).
\end{equation}
Since
\begin{eqnarray}
p_{1}(x_{1}, x_{2}, \gamma) &=& x_{1}\cosh{\gamma} + x_{2}\sinh{\gamma} \nonumber \\
&=& x_{1}\cosh{\gamma} + \left(\frac{a+c}{2} - x_{1}\cosh{\gamma}\right)\mathrm{csch}{\gamma}\sinh{\gamma}\nonumber\\ &=&\frac{a+c}{2}
\end{eqnarray}
it follows that 
\begin{equation}
u^{Q}_{1}(x_{1}, x_{2}) = \frac{1}{4}(a-c)^2.
\end{equation}
%%%%%%%%%%%%%%%%%%%%%%%%%%%%%%%%%%%%%%
Similarly, the set of Nash equilibrium profiles
\begin{equation}
\left\{(0,x_{2}), x_{2} \in \left(\frac{a+c}{2\sinh{\gamma}}, \frac{a}{\sinh{\gamma}}\right)\right\}
\end{equation}
is more profitable for player 1. Since $p_{1}(x_{1},x_{2}, \gamma) = x_{2}\sinh{\gamma}$ and $x_{2} < a/\sinh{\gamma}$, we have $p_{1}(x_{1},x_{2}, \gamma) < a$. Hence,
\begin{equation}
u^{Q}_{1}(x_{1},x_{2}) = (x_{2}\sinh{\gamma} - c)(a-x_{2}\sinh{\gamma}). 
\end{equation}
Thus, there are continuum many nonclassical equilibria that favor player 1. Another part of equilibrium profiles favors player 2 (see Table 1). Moreover, the great part of the equilibria implies payoff of 0 for each player.
%%%%%%%%%%%%%%%%%%%%%
\begin{table}
\caption{Nash equilibria in the game determined by~(\ref{quantumu}) and (\ref{quantumu2}) for $\gamma \ne 0$. They correspond to the points of intersection of graphs of (\ref{bestreplyq1}) and (\ref{bestreplyq2}).}
\label{table1}       % Give a unique label
\begin{tabular}{ll}
\hline\noalign{\smallskip}
Nash equilibrium&Payoff profile  \\
\noalign{\smallskip}\hline\noalign{\smallskip}
$\left(\frac{c}{\mathrm{e}^{\gamma}}, \frac{c}{\mathrm{e}^{\gamma}}\right)$&$(0,0)$ \\
$\left\{\left(x_{1}, \left(\frac{a+c}{2}-x_{1}\cosh{\gamma}\right)\mathrm{csch}{\gamma}\right), x_{1}\in \left[0,\frac{c}{\mathrm{e}^{\gamma}}\right]\right\}$ &$\left(\frac{1}{4}(a-c)^2,0\right)$\\
$\left\{\left(\left(\frac{a+c}{2} - x_{2}\cosh{\gamma}\right)\mathrm{csch}{\gamma}, x_{2} \right), x_{2} \in \left[0,\frac{c}{\mathrm{e}^{\gamma}}\right]\right\}$&$\left(0,\frac{1}{4}(a-c)^2\right)$\\
$\left\{(0,x_{2}), x_{2} \in \left(\frac{a+c}{2\sinh{\gamma}}, \frac{a}{\sinh{\gamma}}\right)\right\}$ & $\left(\left(x_{2}\sinh{\gamma} - c\right)\left(a-x_{2}\sinh{\gamma}\right),0\right)$\\
$\left\{(x_{1},0), x_{1} \in \left(\frac{a+c}{2\sinh{\gamma}}, \frac{a}{\sinh{\gamma}}\right)\right\}$ & $\left(0,\left(x_{1}\sinh{\gamma} - c\right)\left(a-x_{1}\sinh{\gamma}\right)\right)$\\
$\left\{(x_{1},x_{2}), x_{1} \in \left[0, \frac{c}{\mathrm{e}^{\gamma}}\right], x_{2} \in \left[\frac{a}{\sinh{\gamma}}, \infty\right)\right\}$& $(0,0)$\\ $\left\{(x_{1},x_{2}), x_{1} \in \left[\frac{a}{\sinh{\gamma}}, \infty\right),x_{2} \in \left[0, \frac{c}{\mathrm{e}^{\gamma}}\right]\right\}$& $(0,0)$\\ $\left\{(x_{1},x_{2}), x_{1} \in \left[\frac{a}{\sinh{\gamma}}, \infty\right), x_{2} \in \left[\frac{a}{\sinh{\gamma}}, \infty\right)\right\}$ & $(0,0)$\\
\noalign{\smallskip}\hline
\end{tabular}
\end{table}
%%%%%%%%%%%%%%%%%%%%%%%%%%%%
%%%%%%%%%%%%%%%%%%%%%%%%%%%%%%%%%%%%%
\subsection{Bertrand duopoly with fully correlated quantities}
We have known from the previous subsection that the Li--Du--Massar approach to the Bertrand duopoly does not bring us closer to the unique and paretooptimal outcome. It results from similar structure of payoff functions (\ref{quantumu}), (\ref{quantumu2}) and (\ref{bertrand1}), (\ref{bertrand2}). In both cases, a unilateral (slight) deviation from profile $(x,x)$ may yield the player almost twice as high payoff. In this way, we cannot obtain a symmetric equilibrium that would be profitable for both players. However, the correlation between prices can be defined in many different ways. In paper \cite{remarks}, we introduced a simplified model that correlate the players choices $x_{1}, x_{2}\in [0,\infty)$ in the following way:
\begin{equation}\label{mojep}
p'_{1}(x_{1},x_{2},\gamma) = x_{1}\cos^2{\gamma} + x_{2}\sin^2{\gamma}, ~~p'_{2}(x_{1},x_{2},\gamma) = x_{2}\cos^2{\gamma} + x_{1}\sin^2{\gamma}, ~~\gamma \in \left[0,\frac{\pi}{4}\right].
\end{equation}
The value $p'_{i}$ for $i=1,2$ is obtained by formula $p'_{i} = \mathrm{tr}{(M_{i}(x_{1}, x_{2})\rho_{i})}$ where
\begin{enumerate}
\item $M_{i}(x_{1},x_{2}) = \begin{cases}x_{1}|0\rangle \langle 0| + x_{2}|1\rangle \langle 1| &\mbox{if}~ i=1,\cr x_{2}|0\rangle \langle 0| + x_{1}|1\rangle \langle 1| &\mbox{if}~ i=2,\end{cases}$ \item $\rho_{1}$ and $\rho_{2}$ are the reduced density operators $\mathrm{tr}_{2}{(|\Psi\rangle \langle \Psi|)}$ and $\mathrm{tr}_{1}{(|\Psi\rangle \langle \Psi|)}$, respectively, and $|\Psi\rangle = \cos{\gamma}|00\rangle + i\sin{\gamma}|11\rangle$.
\end{enumerate}
It is clear from~(\ref{mojep}) that the correlation between $x_{1}$ and $x_{2}$ increases with increasing parameter~$\gamma$. If $\gamma = 0$, we obtain the classical Betrand duopoly. In the maximally entangled case, $\gamma = \pi/4$, we have $p'_{1} = p'_{2} = p'$ and $p'$ equally depends on $x_{1}$ and $x_{2}$. We will show below that this case is crucial in obtaining paretooptimal equilibria.
\subsubsection*{Nash equilibrium analysis}
Let us first consider the case $\gamma \in (0, \pi/4)$. Substituting~(\ref{mojep}) into (\ref{bertrand1}) and (\ref{bertrand2}) gives
\begin{eqnarray}\label{moju}
u^{Q}_{1}(x_{1},x_{2}) = \begin{cases}\left(p'_{1}(x_{1},x_{2},\gamma)-c\right)\left(a-p'_{1}(x_{1},x_{2},\gamma)\right) &\mbox{if} x_{1}<x_{2}, p_{1}(x_{1},x_{2},\gamma) \leqslant a, \cr \frac{1}{2}\left(p'_{1}(x_{1},x_{2},\gamma)-c\right)\left(a-p'_{1}(x_{1},x_{2},\gamma)\right) &\mbox{if}~ x_{1} = x_{2}, x_{1}\leqslant a, \cr 0 &\mbox{if}~ otherwise,\end{cases} \\ \label{moju2}
u^{Q}_{2}(x_{1},x_{2}) = \begin{cases}\left(p'_{2}(x_{1},x_{2},\gamma)-c\right)\left(a-p'_{2}(x_{1},x_{2},\gamma)\right) &\mbox{if}~ x_{2}<x_{1}, p_{2}(x_{1},x_{2},\gamma) \leqslant a, \cr \frac{1}{2}\left(p'_{2}(x_{1},x_{2},\gamma)-c\right)\left(a-p'_{2}(x_{1},x_{2},\gamma)\right) &\mbox{if}~ x_{1} = x_{2} and x_{2}\leqslant a, \cr 0 &\mbox{if}~ otherwise,\end{cases}
\end{eqnarray}
where we use the fact that inequalities $p'_{1} < p'_{2}$ and $x_{1} < x_{2}$ are equivalent for $\gamma \in (0, \pi/4)$.
We see that functions~(\ref{moju}) and (\ref{moju2}) have the same form as (\ref{quantumu}) and (\ref{quantumu2}) up to the values $p_{i}$. Hence, the method to find $\beta_{1}(x_{2})$ and $\beta_{2}(x_{1})$ and then the Nash equilibria is similar to that used for~(\ref{bestreplyq1}) and (\ref{bestreplyq2}). We obtain
\begin{equation}\label{x1}
\beta_{1}(x_{2}) = \begin{cases}\{x_{1}|x_{1}>x_{2}\} &\mbox{if}~ x_{2} <c,\cr \{x_{1}|x_{1} \geqslant c\} &\mbox{if}~ x_{2} = c, \cr \varnothing &\mbox{if}~ c< x_{2} \leqslant \frac{a+c}{2}, \cr \left(\frac{a+c}{2} - x_{2}\sin^2{\gamma}\right)\cos^{-2}{\gamma} &\mbox{if}~ \frac{a+c}{2} < x_{2} \leqslant \frac{a+c}{2\sin^2{\gamma}}, \cr 0 &\mbox{if}~ \frac{a+c}{2\sin^2{\gamma}} < x_{2} < \frac{a}{\sin^2{\gamma}}, \cr [0,\infty) &\mbox{if}~ x_{2} \geqslant \frac{a}{\sin^2{\gamma}}.\end{cases}
\end{equation}
Symmetric arguments apply to $\beta_{2}(x_{1})$,
\begin{equation}\label{x2}
\beta_{2}(x_{1}) = \begin{cases}\{x_{2}|x_{2}>x_{1}\} &\mbox{if}~ x_{1} <c,\cr \{x_{2}|x_{2} \geqslant c\} &\mbox{if}~ x_{1} = c, \cr \varnothing &\mbox{if}~ c< x_{1} \leqslant \frac{a+c}{2}, \cr \left(\frac{a+c}{2} - x_{1}\sin^2{\gamma}\right)\cos^{-2}{\gamma} &\mbox{if}~ \frac{a+c}{2} < x_{1} \leqslant \frac{a+c}{2\sin^2{\gamma}}, \cr 0 &\mbox{if}~ \frac{a+c}{2\sin^2{\gamma}} < x_{1} < \frac{a}{\sin^2{\gamma}}, \cr [0,\infty) &\mbox{if}~ x_{1} \geqslant \frac{a}{\sin^2{\gamma}}.\end{cases}
\end{equation}
The resulting Nash equilibria in the game are given in table~\ref{table2}.
\begin{table}
\caption{Nash equilibria in the game determined by~(\ref{moju}) and (\ref{moju2}) for $\gamma \in (0, \pi/4)$. They correspond to the points of intersection of (\ref{x1}) and (\ref{x2}).}
\label{table2}       % Give a unique label
\begin{tabular}{ll}
\hline\noalign{\smallskip}
Nash equilibrium&Payoff profile  \\
\noalign{\smallskip}\hline\noalign{\smallskip}
$(c, c)$&$(0,0)$\\
$\left\{\left(x_{1}, \left(\frac{a+c}{2}-x_{1}\cos^2{\gamma}\right)\sin^{-2}{\gamma}\right), x_{1}\in [0,c]\right\}$ &$\left(\frac{1}{4}(a-c)^2,0\right)$\\ $\left\{\left(\left(\frac{a+c}{2} - x_{2}\cos^2{\gamma}\right)\sin^{-2}{\gamma}, x_{2} \right), x_{2} \in [0,c]\right\}$&$\left(0,\frac{1}{4}(a-c)^2\right)$\\ $\left\{(0,x_{2}), x_{2} \in \left(\frac{a+c}{2\sin^2{\gamma}}, \frac{a}{\sin^2{\gamma}}\right)\right\}$ & $\left(\left(x_{2}\sin^2{\gamma} - c\right)\left(a-x_{2}\sin^2{\gamma}\right),0\right)$\\ $\left\{(x_{1},0), x_{1} \in \left(\frac{a+c}{2\sin^2{\gamma}}, \frac{a}{\sin^2{\gamma}}\right)\right\}$ & $\left(0,\left(x_{1}\sin^2{\gamma} - c\right)\left(a-x_{1}\sin^2{\gamma}\right)\right)$\\ $\left\{(x_{1},x_{2}), x_{1} \in \left[0, c\right], x_{2} \in \left[\frac{a}{\sin^2{\gamma}}, \infty\right)\right\}$& $(0,0)$\\ $\left\{(x_{1},x_{2}), x_{1} \in \left[\frac{a}{\sin^2{\gamma}}, \infty\right),x_{2} \in [0, c]\right\}$& $(0,0)$\\ $\left\{(x_{1},x_{2}), x_{1} \in \left[\frac{a}{\sin^2{\gamma}}, \infty\right), x_{2} \in \left[\frac{a}{\sin^2{\gamma}}, \infty\right)\right\}$ & $(0,0)$\\
\noalign{\smallskip}\hline
\end{tabular}
\end{table}
Comparison of Tables 1 and 2 shows that there is no new type of Nash equilibria in the game defined by (\ref{moju}) and (\ref{moju2}).

The set of Nash equilibria changes if $\gamma = \pi/4$. For any $x_{1}$ and $x_{2}$ we have
\begin{equation}
p'_{1}(x_{1}, x_{2}, \pi/4) = p'_{2}(x_{1}, x_{2}, \pi/4) = (x_{1} + x_{2})/2.
\end{equation}
Substituting $p'_{i}(x_{1},x_{2}, \pi/4)$ for $i=1,2$ into (\ref{bertrand1}) and (\ref{bertrand2}), respectively, we can rewrite~(\ref{moju}) and (\ref{moju2}) as
\begin{equation}\label{mojostatni}
u^{Q}_{1}(x_{1}, x_{2}) = u^{Q}_{2}(x_{1},x_{2}) = \begin{cases}\frac{1}{2}\left(\frac{x_{1} + x_{2}}{2} - c\right)\left(a - \frac{x_{1}+x_{2}}{2}\right) &\mbox{if}~ x_{1}+x_{2} \leqslant 2a \cr 0 &\mbox{if}~ x_{1} + x_{2} >2a.\end{cases}
\end{equation}
Note that function $u^{Q}_{1}(x_{1}, x_{2})$ attains its maximum at $x_{1} = a+ c - x_{2}$ for fixed $x_{2} \in [0, 2a)$. Player 1's best reply is therefore $x_{1} = a+c - x_{2}$ ~if~ $0 \leqslant x_{2} \leqslant a+c$ and $x_{1} = 0$ for case $a+c < x_{2} < 2a$. If $x_{2}\geqslant 2a$, then for any $x_{1} \in [0,\infty)$ player 1's payoff is zero. As a result, player 1's best reply function $\beta_{1}(x_{2})$  is given by formula
\begin{eqnarray}\label{b1}
\beta_{1}(x_{2}) = \begin{cases}a+c-x_{2} &\mbox{if}~ x_{2}\leqslant a+c, \cr 0 &if a+c <x_{2} < 2a, \cr [0,\infty) &\mbox{if}~ x_{2} \geqslant 2a.\end{cases}
\end{eqnarray}
\begin{figure}[t]
\centering
\includegraphics[scale=0.7]{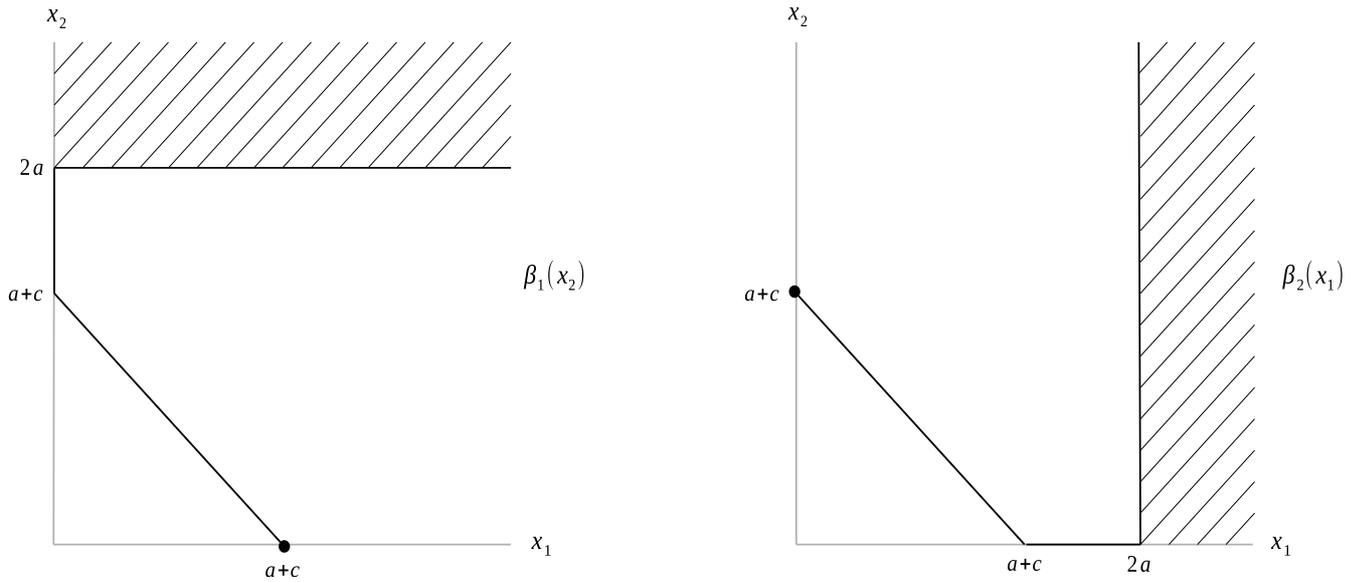}
%\resizebox*{0.72\textwidth}{0.36\textheight}{\includegraphics[scale=1]{figpay6_111_w_ghz.jpg}}
\caption{Graphs of best reply functions (\ref{b1}) and (\ref{b2}).\label{figure2}}
\end{figure}
We conclude similarly that player 2's best reply function $\beta_{2}(x_{1})$ is of the form
\begin{eqnarray}\label{b2}
\beta_{2}(x_{1}) = \begin{cases} a+c-x_{1} &\mbox{if}~ x_{1}\leqslant a+c, \cr 0 &\mbox{if}~ a+c <x_{1} < 2a, \cr [0,\infty) &\mbox{if}~ x_{1} \geqslant 2a.\end{cases}
\end{eqnarray}
From intersection of (\ref{b1}) and (\ref{b2}) (see Fig~\ref{figure2}), we conclude that there are two types of Nash equilibria. One of the types is characterized by profiles $(x_{1},x_{2})$ such that $x_{1},x_{2} \in [2a,\infty)$. In this case, each player's payoff is zero. The other one is what is desired. Each strategy profile $(x_{1},x_{2})$, where $x_{1} + x_{2} = a+c$, constitutes a Nash equilibrium and implies payoff $(a-c)^2/8$ for both players. The payoff profile $((a-c)^2/8, (a-c)^2/8)$ is paretooptimal since
\begin{equation}
\max_{p_{1}, p_{2} \in [0,\infty)}(u_{1}(p_{1}, p_{2}) + u_{2}(p_{1}, p_{2})) = \frac{(a-c)^2}{4},
\end{equation}
where $u_{i}(p_{1},p_{2})$ for $i=1,2$ are the payoff functions (\ref{bertrand1}) and (\ref{bertrand2}).
\begin{table}
\caption{Nash equilibria in the game determined by~(\ref{mojostatni}).}
\label{table3}       % Give a unique label
\begin{tabular}{ll}
\hline\noalign{\smallskip}
Nash equilibrium&Payoff profile  \\
\noalign{\smallskip}\hline\noalign{\smallskip}
$\left\{(x_{1}, x_{2}), x_{1},x_{2} \in [2a,\infty)\right\}$&$(0,0)$\\
$\left\{(x_{1}, a+c-x_{1}), x_{1}\in [0,a + c]\right\}$&$((a-c)^2/8,(a-c)^2/8)$\\
\noalign{\smallskip}\hline
\end{tabular}
\end{table}
\section{Conclusions}
Our research has shown that the quantum approach to the Bertrand duopoly exhibits Nash equilibria that are not available in the classical game. It has been proved that the Li--Du--Massar scheme does not imply the unique and paretooptimal equilibrium outcome as is shown in the Cournot duopoly example. Instead, we have the two equivalent types of Nash equilibria where, depending on the type, one of the players obtains payoff $(a-c)^2/4$ and the other one obtains zero payoff. On the other hand, we have shown that it is possible to attain the symmetric paretooptimal equilibrium if the correlation between the players' prices is defined in another way. If the players share the maximally entangled two-qubit state then with the appropriately defined quantum scheme we obtain the symmetric and paretooptimal equilibrium outcome $((a-c)^2/8, (a-c)^2/8)$. Bertand doupoly, even though it is regarded as unrealistic, attracted a lot of attention in the economic literature. We have shown that "quantization" of the model substantially extends the class of possible player behaviors and Nash equilibria. The above-mentioned paretooptimal equilibrium outcome is exceptionally interesting. We do not claim that there are any quantum correlations among the agents. But, from the phenomenological point of view, agent behaves as if some "coordination" really occurs. This probably means that there are  big incentives to cooperate (collusion). We refrain from speculation on the causes of such behavior \cite{bus,hkh}. Our discussion is yet another argument for using the formalism of quantum  games in the analysis of the  oligopoly modeling.
\section*{Acknowledgments}
\noindent We would like to thank the Reviewers for discussion which undoubtedly improved the quality of our work.\\
\noindent Work by Piotr Fr\c{a}ckiewicz was supported by the Ministry of Science and Higher Education under the project Iuventus Plus IP2014 010973.
%%%%%%%%%%%%%%%%%%%%%%%%%%%%%%%%%%%%%%

\end{document}